# Revisiting LTE LAA: Channel Access, QoS, and Coexistence with Wi-Fi

Jacek Wszołek, Szymon Ludyga, Wojciech Anzel, and Szymon Szott

*Abstract—* Network operators are looking towards LTE License Assisted Access (LAA) as a means of extending capacity by offloading traffic to unlicensed bands. However, operation in these bands requires abiding to certain coexistence rules in terms of channel access. The description of these rules in existing literature is not always in line with the latest standards. Therefore, in this paper, we clarify the operation of LAA, focusing on channel access and methods of providing Quality of Service (QoS) support. In terms of coexistence, we evaluate the impact of LAA under its various QoS settings on Wi-Fi performance in an experimental testbed. Finally, we describe the upcoming research challenges for LTE-based technologies in unlicensed bands considering the latest developments.

## I. Introduction

OFFLOADING data traffic into unlicensed bands is a method considered by 3GPP to alleviate LTE's capacity problems. However, national regulators want to ensure that fair channel access is given to all technologies operating in unlicensed bands. This has led to the studies of coexistence between LTE and Wi-Fi, the predominant technology operating in the aforementioned bands [1].

The use of unlicensed spectrum was initially proposed by the LTE-Unlicensed (LTE-U) Forum to showcase how to leverage unlicensed frequencies in an LTE-like manner. However, because LTE-U did not implement a listen-before-talk (LBT) mechanism, its commercial deployments were possible only in countries in which regulators did not require such a mechanism, such as in the US and China. License Assisted Access (LAA), as the successor of LTE-U, became standardized by the 3GPP in 2015 with Release 13. The main difference between LTE-U and LAA lies in channel access: using LBT allows LTE-LAA to co-exist with incumbent access technologies such as Wi-Fi on a "fair" and "friendly" basis [2][3].

The main goal of LAA is to enable offloading data traffic into unlicensed 5 GHz bands. This is achieved through LTE's *carrier aggregation* and *supplemental downlink* protocols. In principle, LTE offers better coverage and higher spectral efficiency compared to Wi-Fi and, with LAA, allows seamless flow of data across licensed and unlicensed bands in a single core network. From the user perspective, this should translate to an enhanced broadband experience, higher data rates, seamless use of both licensed and unlicensed bands, with high reliability and robust mobility through the licensed anchor carrier.

The study of coexistence issues between LTE and Wi-Fi has been the subject of recent research (Table 1). We have found that while many papers cover LAA (as the standardized successor to LTE-U), few discuss the exact operation of its channel access mechanism (LBT) and usually present results only from analytical or simulation models. Additionally, LBT itself has changed in recent years. European regulations can be satisfied by adherence to LBT requirements that are formally defined in ETSI EN 301 893, and its latest version is 2.1.1 (published in May 2017, effective since June 2018). The 3GPP specifications defining LAA, which largely satisfy the requirements of EN 301 893 in terms of channel access, have likewise been evolving. However, many research and tutorial papers on LAA (Table I) describe channel access according to 3GPP's Technical Report (TR) 36.889 (2015), which references an older version of the ETSI specification, and not as standardized in Technical Specification (TS) 36.213 (2018), which itself has recently been superseded by TS 37.213 (2020).

Based on the above considerations, in this tutorial we briefly describe the operation of LAA (i.e., Licensed-Assisted Access with LTE) focusing on downlink transmissions (the predominantly deployed variant of LAA). We then provide the following contributions: (1) clarifying LBT rules, including a description of the changes introduced in the latest ETSI and 3GPP standards and methods for ensuring Quality of Service (QoS); (2) evaluating the impact of LAA under its various QoS settings on Wi-Fi performance in a standardized experimental testbed; (3) identifying research challenges for 3GPP technologies in unlicensed bands. Finally, we conclude the paper with a prediction of the future of LAA.

## II. Overview of LTE-LAA

LAA is tightly integrated with LTE: it relies on carrier aggregation to combine an LTE carrier from a licensed band (the primary carrier) with an LTE carrier in the 5 GHz

We kindly thank the anonymous reviewers for their insightful comments, which helped improve the paper. The work of J. Wszołek and S. Szott was supported by the Polish Ministry of Science and Higher Education with the subvention funds of the Faculty of Computer Science, Electronics and Telecommunications of AGH University.

J. Wszołek is with AGH University, Krakow, Poland and with Ericsson, Krakow, Poland. S. Ludyga was with AGH University, Krakow, Poland. He is now with Brainhub, Warsaw, Poland. W. Anzel is with Nokia Solutions and Networks, Krakow, Poland (e-mail: wojciech.anzel@nokia.com). S. Szott is with AGH University, Krakow, Poland (e-mail: szott@agh.edu.pl).

TABLE I
COMPARISON OF RECENT STATE OF THE ART IN LTE/WI-FI COEXISTENCE

| Paper | Year | Paper focus | Performance analysis method | Results for LAA | Description of QoS | Latest 3GPP LAA-related specification referenced |
|---|---|---|---|---|---|---|
| [2] | 2017 | LAA, Wi-Fi offloading | Analytical model | Yes | Traffic classes mentioned | TR 36.889 (2015) |
| [3] | 2018 | New LAA-based channel sharing mechanism | Analytical model | Yes | Yes (no reference to QCIs) | TR 36.889 (2015) |
| [4] | 2016 | Carrier sensing for LTE-U | Simulation | No | No | N/A |
| [5] | 2016 | LAA | Simulation | Yes | No | TR 36.889 (2015) |
| [6] | 2016 | LAA | Simulation | Yes | Only VoIP outage | TR 36.889 (2015) |
| [7] | 2017 | LAA | 3GPP results | Yes | No | TR 36.889 (2015) |
| [8] | 2017 | LAA | Simulation | Yes | No | TS 36.213 (2017) |
| [9] | 2018 | LAA | Analytical model | Yes | No | TS 36.213 (2017) |
| [10] | 2018 | LTE-U, LAA | Experiments | Yes | No | TR 36.889 (2015) |
| [11] | 2018 | LAA | Analytical model | Yes | Yes | TS 36.213 (2017) |
| [12] | 2019 | LTE-U, LAA | Simulation and experiments | Yes | Yes | TR 36.889 (2015) |
| [13] | 2020 | LAA | Analytical model, simulations | Yes | Yes | TR 36.889 (2015) |
| **This tutorial** | **2020** | **LAA** | **Experiments** | **Yes** | **Yes** | **TS 37.213 (2020)** |

unlicensed band (the secondary carrier). The primary carrier ensures reliable control signaling and robust, real-time user data transmission with LTE's QoS assurance while the secondary carrier provides data speed bursts. Thus, end users experience both LTE's reliable performance and additional data speed bursts through the unlicensed band. Given the amount of unlicensed spectrum available, using only a single 20 MHz carrier with 2x2 MIMO (Multiple Input, Multiple Output) and 64 QAM (Quadrature Amplitude Modulation), LTE can provide up to 151 Mb/s (in addition to the throughput achieved on the primary carrier). Assuming a perfect LBT success rate, no retransmissions, and the start and stop of transmissions only at subframe boundaries, we have empirically evaluated the maximum LAA channel occupancy time to be around 89% (a result of combining LBT with LAA's scheduled channel access, as described in the following sections). This leads to an upper bound LAA carrier throughput of 134 Mb/s; higher modulations and more antennas result in even higher throughput (up to 357 Mb/s for 256 QAM and 4x4 MIMO).

At the MAC (Medium Access Control) layer, LAA differs slightly from LTE's operation purely in licensed bands. Recall that medium access in LTE is fully under the control of the base station (the evolved node B, eNB) in terms of both downlink and uplink scheduling. The eNB disseminates scheduling decisions, which include transmit power control commands, downlink assignments or uplink grants. In this architecture, the user equipment (UE) cannot send anything (apart from a service request using the random access procedure) in the uplink without the eNB's permission. To receive such an uplink grant, a UE must either send a resource allocation (scheduling) request or report its buffer status at the request of the eNB.

Similarly to LTE operation in licensed bands, the eNB in LAA retains full control of scheduling on the secondary carrier. This leads to a lack of contention among UEs connected to a single eNB, but, due to the presence of other, non-LAA transmissions, the actual transmissions are subject to coexistence mechanisms (such as LBT).

## III. COEXISTENCE WITH OTHER TECHNOLOGIES

LAA's ability to coexist with other technologies is based on selecting an appropriate channel for communication and then on accessing the channel in a fair manner. With respect to the former, LAA implements dynamic channel selection (DCS) which measures each channel's congestion (in terms of transmissions from other technologies) and moves traffic to less occupied channels. DCS algorithms are vendor-specific but rely on channel measurements such as transmission success rates and received signal strength. Regarding the latter, LAA's channel access procedures for shared (unlicensed) wireless channels are specified in 3GPP TS 37.213 which largely follows the coexistence rules defined in ETSI EN 301 893. We proceed with a description of the LBT rules, how they have evolved, and specifics regarding their implementation in LAA.

### A. Principles of LBT

The channel access rules of LAA are based on an LBT mechanism with prioritized, truncated exponential backoff:

1. A device initiating a transmission first waits for the channel to be idle for 16 μs. An idle channel is when there is no other transmission detected above an energy detection threshold level between -75 and -85 dBm/MHz (depending on the maximum transmit power of the coexisting device).
2. Next, the device performs a clear channel assessment (CCA) after each of the $m$ required observation slots (each slot lasting at least 9 μs). A successful CCA decrements $m$ by 1, whereas channel occupancy resets $m$. Once $m$ reaches 0, the device can proceed to the backoff stage.
3. For the backoff stage, the device selects a random integer $N$ in $\{0, ..., CW\}$, where $CW$ is the contention window. CCA is performed for each observation slot and results either in decrementing $N$ by 1 or freezing the backoff procedure. Once $N$ reaches 0, a transmission may commence.
4. The length of the transmission is upper bounded by the maximum channel occupancy time (MCOT), explained in the next section, but never longer than 10 ms.

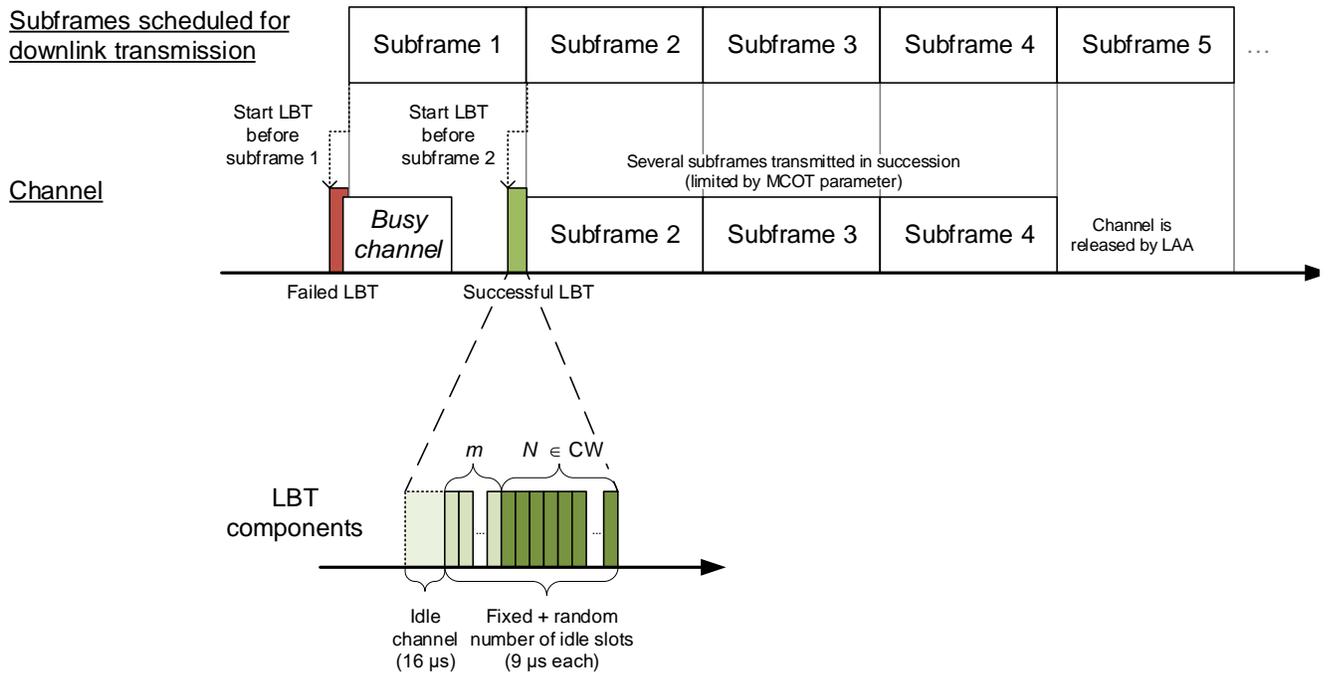

Fig. 1. Channel access in LAA.

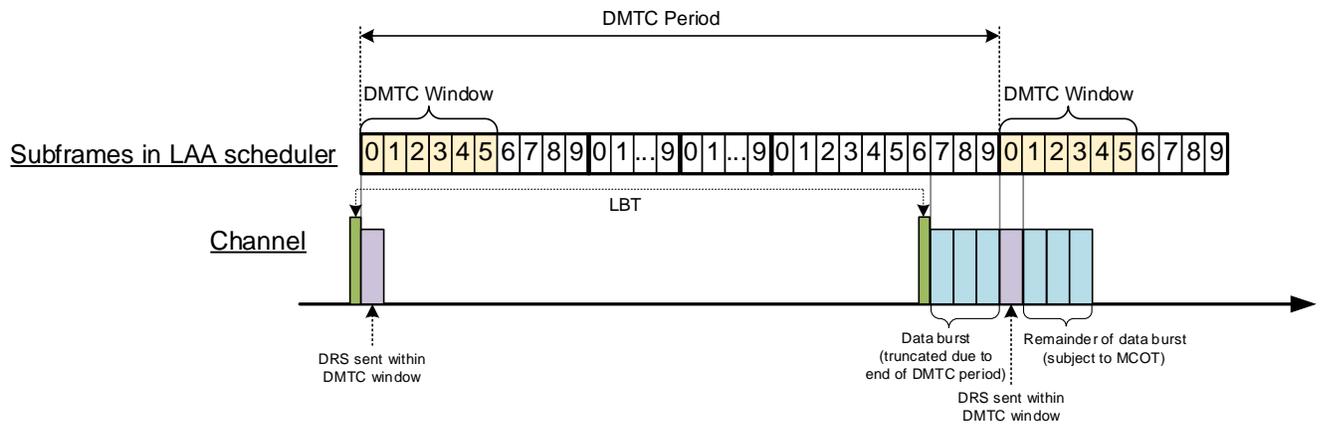

Fig. 2. Example DRS measurement timing configuration (DMTC) behavior for a DMTC period of 40 ms (actual DRS scheduling is vendor-specific).

5. If the transmission is successful, the responding device may send an immediate acknowledgement (without a CCA) and reset *CW* to *CWmin*. If the transmission fails, the *CW* value is doubled (up to *CWmax*) before the next retransmission.

*B. Evolution of LBT*

Before explaining how LBT is used by LAA, we believe it is necessary to explain the changes that occurred between two main versions of LBT: the first was described in ETSI EN 301 893 v1.8.0 (2015), the second was described in ETSI EN 301 893 v2.1.1 (2017) and adopted by 3GPP in TS 36.213 (recently moved to TS 37.213). This distinction is important because key changes were made in LBT and part of the literature (Table I) has become outdated in describing LAA's channel access. The outdated version of LBT differs mostly in the following points:

- The minimum required channel idle time prior to transmission was 20 μs.
- There was no fixed number of additional observation slots ($m = 0$).
- The number of random additional observation slots (*N*) was chosen based on one of two options:
  - Option A: *CWmin* and *CWmax* were fixed at 16 and 1024, respectively, observation slots lasted 18 μs, and MCOT was fixed at 10 ms.
  - Option B: *CWmin* was equal *CWmax* and their values were selected by the manufacturer in the range of 4 to 32, observation slots lasted 20 μs and MCOT was set to 13/32 times the selected *CW* value (in ms).
- The backoff countdown procedure was referred to as extended CCA (ECCA); no such term is used in the current standard, although the principle has remained.
- There was no support for QoS traffic differentiation (as explained in the next section).

Therefore, any descriptions of LAA containing the features listed above can be considered obsolete.

TABLE II
LTE PRIORITY LEVELS AND THEIR MAPPING TO LAA CHANNEL ACCESS PARAMETERS (BASED ON 3GPP TS 37.213 V15.3.0 AND 3GPP TS 23.203 V15.5.0). WI-FI PARAMETERS ADDED FOR COMPARISON.

| LTE | | | | LAA | | | | Wi-Fi | | | |
|---|---|---|---|---|---|---|---|---|---|---|---|
| QCI | Resource Type | Priority Level | Example Services | Priority Class | $m$ | $CW_{min}$, $CW_{max}$ | MCOT [ms] | Access Category | $m$ | $CW_{min}$, $CW_{max}$ | MCOT [ms] |
| 1 | GBR | 2 | Conversational Voice | 1 | 1 | 3, 7 | 2 | Voice | 2 | 3, 7 | 2.080 |
| 3 | | 3 | Real Time Gaming, V2X messages | | | | | | | | |
| 5* | Non-GBR | 1 | IMS Signalling | | | | | | | | |
| 2 | GBR | 4 | Conversational Video (Live Streaming) | 2 | 1 | 7, 15 | 3 | Video | 2 | 7, 15 | 4.096 |
| 7* | Non-GBR | 7 | Voice, Video (Live Streaming), Interactive Gaming | | | | | | | | |
| 6 | | 6 | Video (Buffered Streaming) | 3 | 3 | 15, 63 | 8 or 10** | Best Effort | 3 | 15, 1023 | 2.528 |
| 8 | Non-GBR | 8 | TCP-based (www, email, etc.) | | | | | | | | |
| 9* | | 9 | Best effort | | | | | | | | |

* QCI values used in experiments.
** 3GPP TS 37.213 specifies a default MCOT value of 8, which can be increased to 10 ms if there are no other technologies accessing the channel.

## C. LAA Channel Access

The current version of LAA's LBT resembles the enhanced distributed channel access (EDCA) of IEEE 802.11, which served as a basis for the definition of these rules. However, the main difference in the operation of LAA is that it is based on LTE which employs scheduled channel access, whereas 802.11 uses random channel access. These two opposing methods are united in LAA through the following. As in LTE, the scheduler prepares data to transmit in each 1 ms subframe. This data is prepared with a 4 ms advance and sent to the lower layers of the eNB for transmission. Any UE-decodable data must be sent within a 1 ms duration called the transmission time interval (TTI). The start of the LBT procedure described above needs to be somehow aligned with this mechanism, though how this is exactly done is vendor-specific. One option is an early LBT start and transmission of a reservation signal, which would last from the moment of gaining channel access until the subframe starts. This option allows for early channel reservation but lowers the maximum achievable throughput as the reservation signal is counted into the MCOT and is not favored by 802.11 [14]. Another strategy is to initiate the LBT at an appropriate time prior to the subframe start so that if it is successful, the data transmission can commence immediately. This method provides better throughput results, but one risks that the channel is already reserved, e.g., by another device using the former strategy. Regardless of the method, any data sent from the eNB to the UE must begin at the start of the subframe (though 3GPP allows for minor deviations from this rule, such as Partial Starting and Partial Ending Subframes). LAA uses a hybrid automatic repeat request (HARQ) method to report the success (or failure) of decoding a transmission: the UE sends an acknowledgement (ACK) or negative acknowledgement (NACK) over the licensed band and the scheduler invokes a retransmission procedure which is vendor-dependent but usually has a higher scheduling priority. Also, if the NACKs for a given reference subframe exceed a threshold, the *CW* size is doubled (and reset otherwise) [14].

Fig. 1 illustrates how subframes scheduled for transmission by the LAA scheduler can be sent over an unlicensed channel (assuming no reservation signals are used). For the first subframe the LBT procedure could start but the LAA transmission was preempted by another device accessing the channel. For subframes such as this, which were scheduled but not transmitted, LTE's higher layers will initiate the retransmission of the subframe. Finally, we see that LBT is successful prior to subframe 2 and several subframes may be transmitted in succession (as limited by MCOT). Afterwards, the channel is released.

Additionally, LAA requires a minimum amount of signaling in the unlicensed band for the UE to remain synchronized with the eNB and to estimate the channel. To this end, the eNB sends discovery reference signals (DRS) transmitted in DMTC (DRS measurement timing configuration) windows. A DMTC window always lasts 6 ms (out of which 1 ms is occupied by the DRS). DMTC's periodicity can be set to 40, 80, or 160 ms, according to the operator's preference. If an ongoing data burst occurs during the DMTC, the DRS can be multiplexed with the user data (Fig. 2). If there is no ongoing data burst (but data is scheduled to be sent), a short DRS-only burst will take priority. As we will demonstrate, DMTC's periodicity only slightly impacts LAA's coexistence with Wi-Fi.

## IV. QOS IN LTE-LAA

QoS has become an important part of designing LTE networks for supporting both data and voice services. There are cases in which LTE services are used for critical operations (such as voice calls) and cases where LTE is used in situations where only best effort service is required (such as Internet browsing). To this end, QoS in LTE follows a class-based approach: there are bearers with guaranteed bit rate (GBR) and bearers without such guarantees (Non-GBR). The default bearer, established when the UE attaches to an LTE network (in a licensed channel), is always a Non-GBR bearer. Since LAA cannot provide a guaranteed bit rate, due to contention with other systems, in principle, only Non-GBR bearers can be used. However, the eNB constantly monitors the quality of the unlicensed channels and may attempt to deliver QoS sensitive applications over the unlicensed band if the quality of the channel is deemed sufficient. Nonetheless, the eNB can always revert to the licensed bands (without terminating the connection).

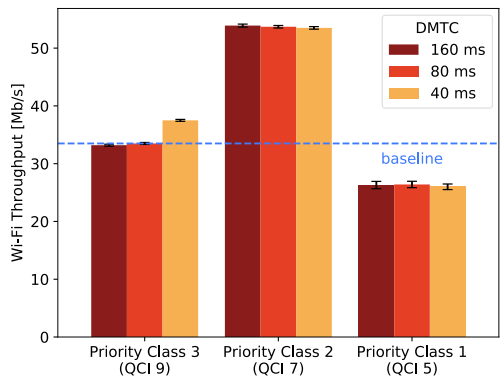

Fig. 3. Wi-Fi throughput performance in the coexistence scenario. The 95% confidence intervals are shown.

At the next level of differentiation, QoS Class Identifiers (QCIs) are used for traffic prioritization. Each QCI is mapped to a priority level (Table II) and if congestion occurs, the lowest priority level traffic is the first to be discarded. Non-GBR bearers (available in LAA) have a QCI from 5 to 9. With subsequent LTE releases, 3GPP has extended the set of QCIs, but none of them are to be used in unlicensed bands.

LTE's QCIs and priority levels are mapped to LAA through the priority classes (Table II). Since channel access is determined by the required number of fixed ($m$) and random ($N$ in $\{0, \ldots, CW\}$ for $CW$ between $CWmin$ and $CWmax$) idle channel slots, classes with higher priority have lower $m$ and lower $\{CWmin, CWmax\}$ values (similar to Wi-Fi's EDCA parameter set). This higher priority in channel access is offset by the maximum transmission length (MCOT), where the lowest priority classes can transmit the longest in a single burst. This reflects both the short packet sizes of high-priority services as well as the bursty nature of best effort traffic. Next, we demonstrate the impact of these priority class settings on coexistence with Wi-Fi.

## V. EXPERIMENTAL RESULTS

We used an experimental test setup, fully compliant with the setup described in ETSI EN 301 893, to measure network performance in an LAA and Wi-Fi coexistence scenario consisting of an eNB, a UE (the transmitter and receiver of LTE traffic), and a commercially available off-the-shelf Wi-Fi access point (AP) with an attached Wi-Fi client.

The Wi-Fi devices operated using default IEEE 802.11ac settings. An attenuator between the eNB and UE was used to protect the radio receivers in both devices against excessive signal power. The eNB used two remote radio heads (RRHs), for the licensed and unlicensed bands. DCS, used for avoiding congested channels, was disabled on the eNB to force LAA to work on the same channel as Wi-Fi. Also, the eNB used reservation signals in LAA to maintain channel access until the beginning of the subframe. Both RRHs transmitted with a power of 16 dBm, whereas the AP worked on the same unlicensed channel with a transceiver power of 20 dBm. At the recipient side, the UE was configured to support LAA, carrier aggregation, and both frequency bands. The experiments used simultaneous LTE and Wi-Fi transmissions, both with full-buffer UDP downlink models of traffic.

The goal of the experiments was to verify how the QCI of the established LAA bearer and the DMTC period of the LAA cell influence the fairness in resource division in the unlicensed band. To vary the QoS of LAA traffic, we used the following bearers:

- QCI 9 for low priority,
- QCI 7 for higher priority,
- QCI 5 for highest priority traffic.

We also changed the DMTC window settings (40, 80, or 160 ms). The most important performance metrics of Wi-Fi traffic (throughput and delay) were measured and calculated using *iPerf* (configured to saturate the network) and *ping* (under default settings) on the Wi-Fi client side (in separate measurements).

### A. Throughput Performance

We began the coexistence measurements by establishing a baseline: the maximum UDP throughput measured at the Wi-Fi station in the absence of LAA traffic. Assuming that fair coexistence in the unlicensed band would result in dividing the resources equally between transmitting devices, the coexistence baseline for Wi-Fi should be 50% of this measured throughput. In our testbed the measured Wi-Fi throughput without LAA was 67 Mbit/s, resulting in a 50% baseline of 33.5 Mbit/s.

Next, we measured Wi-Fi throughput under coexistence with various LAA configurations (Fig. 3). Based on the received results it can be observed that when the LAA cell uses QCI 9 bearers, Wi-Fi throughput is close or slightly above the baseline. This translates to fair, indeed almost perfect, LAA and Wi-Fi coexistence. Furthermore, using a shorter DMTC period (40 ms) resulted in higher Wi-Fi throughput because sending DRS bursts with a higher periodicity, causes the long (up to 8 ms, cf. Table II) data bursts to be terminated or prevented more often than with lower DMTC periodicity.

For QCI 7, LAA's MCOT is decreased almost three times which led to a much higher Wi-Fi throughput (exceeding the baseline by 60%). The reduction of $m$ and $CW$ did not compensate for the loss in transmission length. Also, the reduced MCOT did not allow to observe any impact of DMTC periodicity on Wi-Fi throughput. In this case, LAA is clearly a better neighbor for Wi-Fi.

Finally, when LAA uses QCI 5, the Wi-Fi throughput is around 22% below the baseline. This is caused by the significantly reduced $m$ and $CW$ parameters: assuming a low collision probability, the average backoff slots before transmission for LAA would be 2.5 (the sum of $m$ and half of $CWmin$) as compared to Wi-Fi's 10.5. This leads to the conclusion that in this configuration the level of fairness in terms of coexistence between LTE and Wi-Fi is lower than in cases with QCI 7 and QCI 9 bearers. However, QCI 5 is designed only for IMS signaling and there are throughput limitations on the traffic transferred with this priority (typically no more than 256 kb/s). Therefore, in a real-world deployment, high-rate traffic using QCI 5 should never occur. Meanwhile, QCIs 1 and 3, also from priority class 1, are GBR and thus not eligible for transmission over LAA.

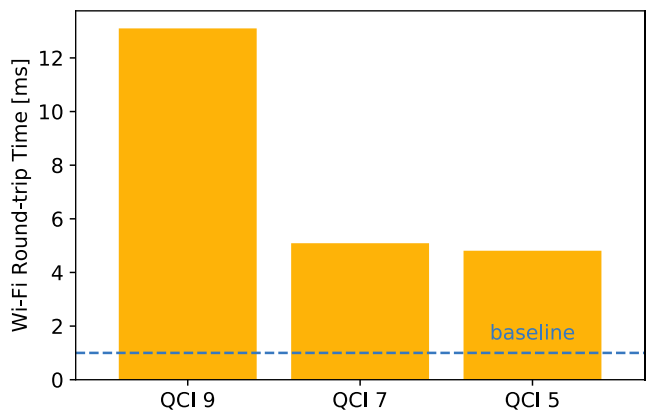

Fig. 4. Wi-Fi delay performance in the coexistence scenario.

*B. Delay Performance*

We analyzed the impact of LAA on the delay of Wi-Fi transmissions by using the round-trip time (RTT) as a metric of latency. We first measured the baseline: the average RTT observed at a Wi-Fi station in the absence of LAA traffic (1 ms in our case).

The results show that LAA traffic has a strong, negative impact on Wi-Fi delay (Fig. 4). In contrast to throughput, the most significant delay degradation occurred when LAA used QCI 9. The observed average RTT was 13.1 ms. This significant increase of RTT for Wi-Fi is caused mostly by the long MCOT time (8 ms) of QCI 9 traffic. Since LAA occupies the channel for 8 ms it must introduce at least 8 ms of additional delay. This explains why the average RTT is higher than 8 ms.

When LAA traffic uses QCI 7, the measured average RTT for Wi-Fi was 5.09 ms. This means an improvement of almost 8 ms compared to priority class 3 LAA traffic, mainly caused by the shorter MCOT (3 instead of 8 ms). A similar trend can be observed when LAA uses QCI 5. The average RTT equals 4.81 ms and is slightly shorter than for priority class 2. We conclude that even though several LBT parameters determine the probability of channel access, MCOT has the most significant impact. Additional results (not presented here) showed that the impact of DMTC periodicity on RTT is negligible.

## VI. CONCLUSIONS AND RESEARCH PERSPECTIVES

Fortunately for Wi-Fi users, many studies (including ours) suggest that LAA can be a better neighbor for Wi-Fi than Wi-Fi itself. Therefore, standardization efforts have been continuing within 3GPP: Enhanced LAA (eLAA), defined in Rel-14 and Rel-15, enables uplink operations which require two successful LBT procedures: one for grant transmission (downlink) and one for the actual data (uplink). Scheduling and grant delays may have a negative impact on uplink performance so we await the availability of eLAA devices to determine their performance. However, eLAA may not be massively deployed as we foresee operators switching to 5G, also in terms of providing unlicensed access, where further performance improvements can be expected. 3GPP Rel-16 will include a 5G-tailored version of LAA called NR-U (New Radio Unlicensed) [15] and we expect both LAA and NR-U to coexist in future years.

With the congestion of existing unlicensed bands (2.4 and 5 GHz), the opening of new bands by national regulators for unlicensed access would provide more resources for the operation of 3GPP-based technologies. One approach is the reallocation of the 6 GHz band which encompasses 1.2 GHz in the US and 500 MHz in Europe. There are incumbents in this band (mainly fixed/mobile radio links and fixed/mobile satellite services) but undergoing regulatory efforts should ensure that in the coming years the 6 GHz band will become available worldwide, following its recent release in the US.

Considering the above, we can identify the following upcoming research challenges for 3GPP technologies in unlicensed bands:

- Evaluating the coexistence of networks using heterogeneous technologies (Wi-Fi, LAA/eLAA, NR-U) especially in the new millimeter wave frequency bands.
- Evaluating the benefits of upgrading from LAA to NR-U.
- Designing new coexistence methods (e.g., for multichannel access) and evaluating them in actual deployments.

These topics are currently being addressed by researchers and standardization bodies and discussed by the major stakeholders in coexistence workshops.

Evaluating LAA from a practical, mobile network operator perspective, it is especially beneficial for those operators who do not have enough licensed bandwidth to meet their needs and better than Wi-Fi-based alternatives such as LTE-WLAN Aggregation (LWA). For these reasons LAA equipment has been selling particularly well in the USA but also in Italy, Hong Kong, Thailand, Turkey, and Russia (according to the Global mobile Suppliers Association's reports). In Europe, LAA is less popular due to radio restrictions imposed on outdoor deployments. Furthermore, in countries such as Poland, where the lack of licensed bandwidth for end users has only recently been observed, LAA is only now being considered for deployment.

In conclusion to this paper, we state that labelling LAA as a "Wi-Fi killer" is incorrect for several reasons. First, both LAA and Wi-Fi have different use cases, with Wi-Fi's strength being standalone deployments and fewer restrictions. Second, if enough free channels are available, coexistence issues can be minimized. This is further alleviated by the upcoming use of higher frequencies (which leads to lower interference between adjacent cells) and the availability of new bands though at the same time magnified by the use of channel aggregation (the next Wi-Fi release will support 320 MHz channels). Third, according to Cisco's Mobile VNI report from February, 2019 over 50% of Internet traffic will be carried by Wi-Fi devices in 2022. It will take some time to radically alter this statistic. For this to occur, 5G will need to become ubiquitous and NR-U equipment – an affordable alternative to Wi-Fi.

**Jacek Wszolek** received his Ph.D. degree in telecommunications from AGH University in 2014. Currently he is working there as an assistant professor. His research interests are related to applying machine learning in cellular broadband wireless communications. In 2019, he joined Ericsson and works there as a System Developer in the RAN Architecture/Technology Analysis program.

**Szymon Ludyga** received his M. Sc. degree in telecommunications from AGH University in 2018. His research interests are related to web development and web security. Recently, he joined Brainhub and works there as a Full Stack Engineer developing software for financial applications.

**Wojciech Anzel** received his M. Sc. degree in electronics and telecommunications from the AGH University in 2013. His research interests are related to various cellular radio access issues, including unlicensed and New Radio topics. Currently he is working as a Specification Engineer at Nokia Mobile Networks.

**Szymon Szott** received his Ph.D. degree in telecommunications from the AGH University in 2011. Currently he is working there as an associate professor. His professional interests are related to wireless local area networks (channel access, QoS provisioning, security). He is the author of over 70 research papers.